# CyberChair: A Web-Based Groupware Application to Facilitate the Paper Reviewing Process

**Richard van de Stadt**
Department of Computer Science
University of Twente
P.O. Box 217, NL-7500 AE  Enschede, The Netherlands
+31 53 489 3731
stadt@cs.utwente.nl
**- Submitted -**

**ABSTRACT**
In this paper we describe CyberChair, a web-based groupware application that supports the review process for technical contributions to conferences. CyberChair deals with most administrative tasks that are involved in the review process, such as storing author information, abstracts, (camera-ready) papers and reviews. It generates several overviews based on the reviews which support the Program Committee (PC) in selecting the best papers. CyberChair points out conflicting reviews and offers the reviewers means to easily communicate to solve these conflicts. In his paper *Identify the Champion* [9], O. Nierstrasz describes this review process in terms of a pattern language. CyberChair supports PCs by using these patterns in its implementation.

**INTRODUCTION**
The review process for contributions for conferences can be rather tedious because of the administrative tasks that come with the process. Up until its 1996 edition, the European Conference on Object-Oriented Programming (ECOOP) used hard-copy review forms. We decided to automate this and provide electronic review forms, by using the World Wide Web (WWW), since all reviewers were now able to use this emerging technology. The result turned out to be very fruitful. The next years the system was improved and more features, such as uploading of papers, were included in the system to support the process. The current version of CyberChair supports all phases of the review process, including preparation of the proceedings in the electronic format prescribed by the publisher. The version described in this paper was used for ECOOP 2000 and the International Workshop on Software Specification and Design (IWSSD-10). Other conferences have used older versions of CyberChair.

**THE REVIEW PROCESS**
Generally, the review process for technical contributions to conferences is as follows. To enter the review process, authors must send their paper to the Program Committee Chair (PCC). When all submissions have arrived, suitable members of the Program Committee (PC)[1] must be selected to review each paper. The PCC sends the papers and review forms to the reviewers, who must fill in the review forms and send these to the PCC. Typically, the review process ends by organizing a Program Committee meeting, prepared by the PCC, where papers are discussed and accepted or rejected for presentation at the conference. After the meeting, the reviewers' comments are sent back to the authors, so that they can improve their paper. The authors of accepted papers may submit a brushed-up version of their paper, in so-called camera-ready format to the PCC, who will send them, together with the preface and the table of contents to the publisher to have the proceedings printed.

**ADMINSTRATIVE TASKS**
The following administrative tasks can be identified during a typical review process: Author information, abstract and paper submission, paper-to-reviewer assignment, paper distribution, filling in and sending review forms, collecting review forms and ordering them, preparing the PC meeting, returning the reviewers' comments to the authors, camera-ready paper submission and preparation of proceedings.

---

[1] Members of the PC are also called 'reviewers'

When email is used for the submission of papers and reviews, the physical distribution of papers, which is still used for many conferences, is no longer needed. However, the administrative tasks are still necessary .

The tasks mentioned above have been automated by CyberChair. Using the WWW, a commonality these days, the ones involved in the process can collaboratively use CyberChair to store and/or retrieve data from password-protected areas controlled by CyberChair.

**USING CYBERCHAIR**

This section explains how the CyberChair handles the administration and supports its users during the complete reviewing process. This is done by showing the options for the 4 kinds of users of CyberChair: the maintainer, the authors, the PCC and the reviewers.

**The Maintainer**

The maintainer (who might be the same person as the PCC) installs CyberChair. This includes editing the files that contain the templates for the PCC and reviewer information, the file containing the main topics of the conference, and setting up a website. The maintainer also makes sure that overviews are generated at regular intervals. Further, the maintainer may need to convert submitted papers to other formats, or ask the authors to submit the paper in another format. The maintainer should make sure that enough disk space is available to store the papers. After the review process, the maintainer makes sure that the notifications of acceptance and rejections, including the reviewers' remarks, are sent to the authors. Finally, after the camera-ready papers have been submitted, the proceedings must be prepared by the maintainer by entering the conference sessions and the numbers of the papers within those sections in a specific template file. If the publisher allows electronic versions, the maintainer should send or upload the proceedings, for further preparation by the publisher.

**The Authors**

Authors must submit their paper in two steps. In the first step, authors must fill in a webform shown by figures 1 and 2, on which they state their names and contact information, the title of their paper, an abstract of the paper, and indicate which conference topics their paper covers CyberChair assigns a unique identification to the submission and sends a login and password combination for step 2 to the contact person's email address. In step 2, the full paper must be submitted by uploading it. The authors must use the login and password combination sent after phase 1.

The authors may make corrections or changes to the information that was submitted in step 1. After the review process, the authors of submitted papers should submit a camera-ready version of their paper, analogous to step 2.

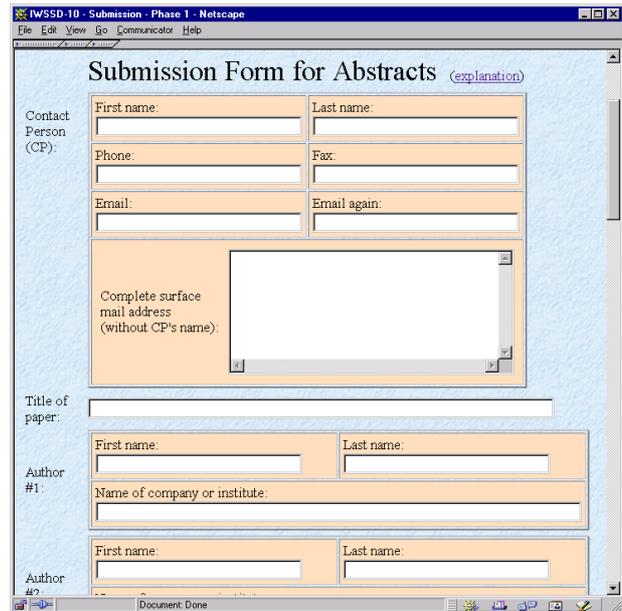

**Figure 1 - Submission form for authors, contact information**

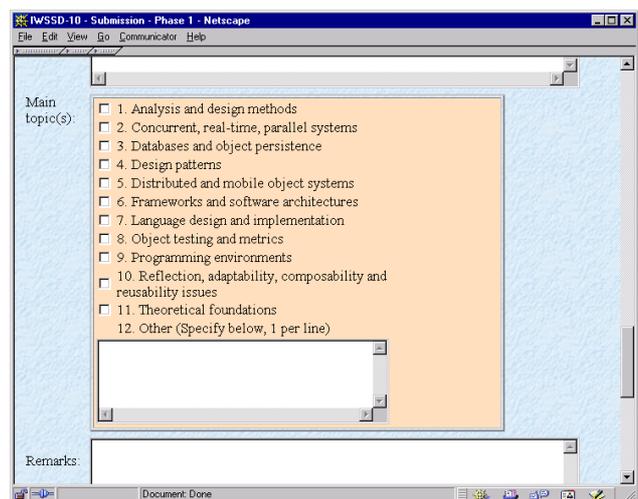

**Figure 2 - Submission form for authors, conference topics**

**The Reviewers**

Identifying the right reviewers for papers is a difficult task. Following the patterns *Experts Review Papers* and *Champions Review Papers* [9], papers should be matched to domain experts in the PC. This can be done by letting the reviewers indicate which papers they would like to review (also called 'bidding'). Papers that have not received enough bids should be assigned based on the expertise of the reviewers.

*Bidding support*

To support reviewers with making choices, CyberChair provides them with several webpages containing overviews of the submitted abstracts. Reviewers can browse the overviews by conference topic, to quickly identify the papers that match their interest and expertise (Figure 3).

Clicking on a topic will show a list of all abstracts of papers that cover that topic, which can then easily be browsed. For each paper that a reviewer would like to review, the corresponding checkbox in the left part of the window should be marked. Further, a high or low priority should be chosen, to indicate how eager they are to review the paper. Reviewers may bid in several stages. The submitted bids are accumulated. Further, overviews are available that contain, per conference topic, all abstracts, and an overview that contains all abstracts. These overviews can be printed in case reviewers want to read the abstracts offline. For this process to work smoothly, the abstracts should be submitted at least one week before the full paper is submitted. Reviewers are also asked to inform the maintainer of papers they do not want to review because of a conflict of interest.

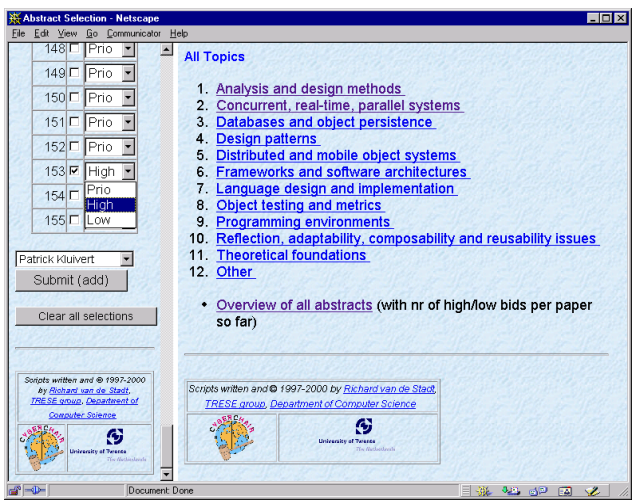

**Figure 3 - Bidding for papers**

*Reviewers' expertise*

The reviewers are asked to indicated their expertise of the conference topics and their willingness to review papers on those topics. This information is used for the paper distribution. Reviewers can indicate whether they are an expert on the topic (X), no expert, but knowledgable in the subject area (Y), or an informed outsider (Z). For the willingness to review papers of certain topics the reviewers can indicate whether they would rather not review such papers (R), or will not review such papers (W). For each topic, either X, Y or Z should be indicated, followed by an optional R or W.

*Reviewers' webpages*

After the paper distribution, CyberChair generates a personal, password-protected webpage for each reviewer. This page consists of several dynamically generated HTML frames. The top frame shows, in colored boxes, the numbers of all papers that have been assigned to the reviewer. The colored boxes indicate the *state* of the reviews, which is determined each time a review is submitted by any of the reviewers. The state is based on the classification given to papers by the reviewers. The rest of the webpage is initially empty.

After clicking on a paper number several hyperlinks are generated to allow the paper to be downloaded, to display the abstract and to submit or update a review. When a reviewer has submitted all his reviews and has time to review more papers, he may look at the overview containing all papers and select papers for which conflicting reviews exist. The reviewer can be added to the list of reviewers for that paper. His or her personal webpage will then be updated automatically.

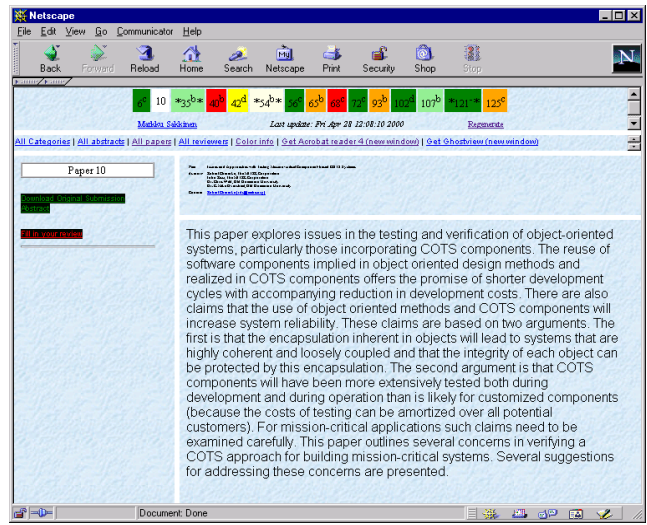

**Figure 4 - A reviewer's page**

In case a reviewer has already submitted a review of the selected paper, hyperlinks to the reviews of the other reviewers are shown, so that the reviewer can read their opinion. In case conflicting reviews exist, the reviewers can use the hyperlink that enables easy communication to the other reviewers by email. A copy of the message is sent to the PCC. Further, links are provided that contain all reviews of the reviewer, and all reviews of all reviewers of the paper, respectively.

The top frame is automatically updated every 5 minutes, to reflect the current state of the reviews of the papers.

*Classifcation of papers*

Following Nierstrasz's pattern *Make Champions Explicit* [9], the reviewers are asked to fill in the classification of papers on the review form. The following classifications are used:

A: "I will champion this paper at the PC meeting (Advocate/Accept)".

B: "I can accept this paper, but I will not champion it (accept, but could reject)".

C: "This paper should be rejected, though I will not fight strongly against it (reject, but could accept)".

D: "Serious problems. I will argue to reject this paper (Detractor)".

In addition, the reviewers must indicate their overall expertise on the topics that a paper covers. This is used to discover papers that have only been reviewed by non-experts, so that the PCC may ask another reviewer to review the paper.

*Conflict detection*

The pattern *Identify the Conflicts* [9] is used to determine the state of the reviews. The highest and lowest classification given to a paper are taken into account. Using a coloring scheme, the state of reviews is indicated. The following colors are used to indicate the state of the reviews on the reviewers' webpages:

- White: You have not yet submitted your review
- Pink: Only your review has been submitted.
- Light green: No classification conflict - A and B only
- Orange: No classification conflict - B and C only
- Green: No classification conflict - C and D only
- Light yellow: Classification conflict - both A and C
- Yellow: Classification conflict - both B and D
- Red: Serious classification conflict - both A and D
- Gold: Accepted Paper

**The PCC**

The PCC's tasks are drastically diminished when CyberChair is used, so that he or she can fully focus on the preparation of the PC meeting. CyberChair can handle the paper distribution, collects and categorizes all reviews and prepares overviews of reviews which can be used during the PC meeting. The table of contents and author index of the proceedings are generated by CyberChair and the camera-ready papers are prepared for electronic delivery to the publisher.

*Paper distribution*

As soon as the full paper deadline has passed, the papers can be assigned to the reviewers. CyberChair generates a paper distribution proposal for the PCC. This is done by combining the reviewers' expertise and willingness to review papers on certain topics with the preferences the reviewers indicated by bidding for papers.

*Monitoring the review process*

The PCC can monitor the review process by using several overviews, which are generated at regular intervals.

- The number of reviews submitted by each reviewer
  This is also available to all reviewers, which might speed up the review process.
- A one-page overview of all reviews and their state, with hyperlinks to the reviews (figure 5).
- An overview with categories of papers, based on their classification (figure 6).
- An overview of 'champions' of papers, i.e. reviewers who indicate they would like these papers to be accepted.
- Low expertise reviews: Papers that have only been reviewed by reviewers who indicated that they are no expert on the topics the paper covers. The chair may ask additional reviewers to read the paper and submit a review.

After the PC meeting, the PCC must inform the maintainer about the accepted papers, so that the notifications can be sent to the authors.

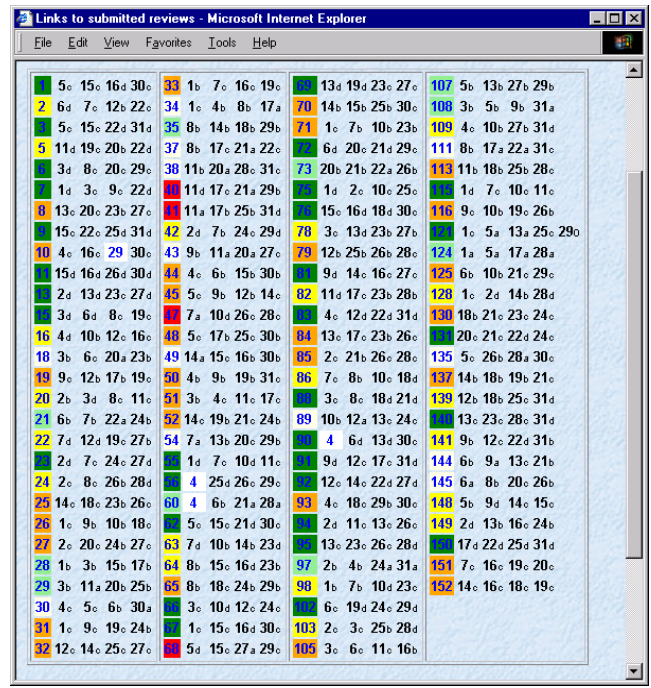

**Figure 5 - Overview of all reviews**

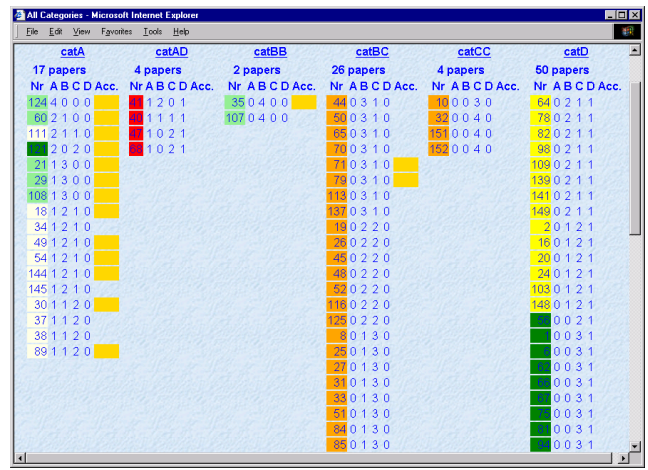

**Figure 6 - Overview of categories**

## IMPLEMENTATION

CyberChair is composed of a set of Common Gateway Interface (CGI) scripts, which are activated by the webforms that are used by the authors and reviewers. Other scripts are so-called administrative scripts which are either started by the maintainer or automatically at regular intervals by the system (e.g. crontab on Unix systems). The CGI scripts have been written in Python, a very powerful, easy-to-learn programming language, which allows rapid implementation. Like Java it runs on a virtual machine and is available for many platforms. The implementation of some parts of CyberChair is not that trivial. Those parts are explained in this section.

### Paper Distribution

The current implementation collects the preferences (bids), expertise levels and willingness to review papers on certain topics and the conflicts of interest of the reviewers.

Currently the script is tuned to assign as much as possible papers to reviewers based on their preferences, provided that all papers are assigned to exactly 4 reviewers and all reviewers are assigned approximately the same number of papers. Graph theory (nodes with labeled edges) is used to solve this assignment problem.

For each paper, a list of the reviewers who indicated a *high* preference for the paper is generated. This list is sorted, based on the number of papers the reviewers have already been assigned so far. The 4 reviewers with the least number of papers assigned so far 'get' the paper. In case there are less than 4 reviewers, this process is repeated with the list of reviewers who indicated a *low* preference for the paper. If there are still not enough reviewers for the paper, the expertise level of reviewers of all topics the paper covers is calculated and the paper is assigned to the reviewers with the highest overall expertise. In case the 'expertise-value' is equal, the reviewer with the least number of papers assigned so far 'gets' the paper. Note that reviewers may have indicated that they do not want to review papers that cover certain conference topics, while they also indicated a preference for certain papers that cover just those topics. The preference 'wins' in such cases.

In case not enough reviewers can be found to review a paper, this is an indication that the script should be better tuned. Apparently the *pool of experts* is not full enough. The size of this pool is influenced by the maximum number of papers that is assigned to reviewers based on their preferences. When a reviewer reaches this maximum, he or she will not be part of the preference lists, because apparently his or her expertise is needed to review 'less popular' papers.

The script is currently optimally tuned for ECOOP 2000. For each conference/workshop the tuning should be considered carefully, since it heavily depends on the number of submissions, the number of reviewers, their expertise and the number of papers per reviewer. Figure 7 shows a part of a paper distribution as generated by CyberChair. In general, either a large percentage of preferences is assigned (in case reviewers selected enough papers that were indeed submitted), or the number of preferences that was assigned is high (in case a reviewer indicated more papers than the maximum number of papers assigned to reviewers).

### Reviewers' pages

The pages of the reviewers are generated once by the maintainer when the paper distribution has been done. As mentioned before, a reviewer's page consists of several frames (a frame set). The top frame, containing the paper numbers in colored boxes, is updated every 5 minutes by using a so-called meta tag containing the 'refresh' option. This allows automatic reloading of a page of which the Universal Resource Location (URL) is indicated in the same meta tag. For the frame to display the current state of reviews, the URL is not the name of a file, but the name of a CGI script which recalculates the state of all papers and 'prints' its output in the frame.

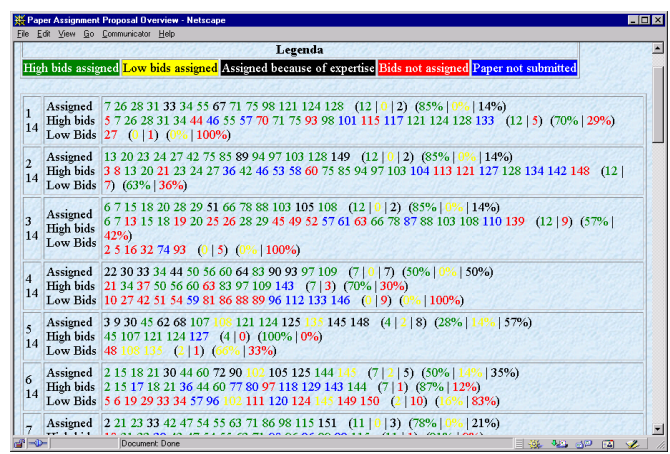

**Figure 7 - Paper distribution overview**

When the reviewer clicks on a paper number, this causes the activation of a CGI script that in its turn generates a frameset, consisting of 3 frames (See figure 4). Each of these frames is filled by using the name of a CGI script in the SRC tag of the frame definition. The left frame will contain a list of hyperlinks for downloading a paper, displaying the abstract, submitting a review, or looking at other reviewers' reviews. The frame in the middle of the browser window will contain the paper title and the names of the authors of the paper. The lower right frame initially contains the abstract of the paper. As soon as the reviewer has submitted his or her review, the review will be the default to be displayed in this frame when the reviewer clicks on the paper number in the top frame.

We are aware that the pages could also have been created by using tables. Sites like Amazon [1] use this style and regenerate a complete new page after every mouse click.

This is quite inefficient, because much of the data that did not change is also regenerated. Using frames allows for updates on specific parts of the screen, so that only data that has changed can be updated without the need to refresh the complete screen. Moreover, since we chose to regenerate the state of all papers displayed on the reviewer's webpage every 5 minutes, this would make it almost impossible for reviewers to enter a review form. Especially that part of the browser window must be left as is while the reviewer is filling in the review form.

After filling in a review form, the reviewer is asked to click on a button labeled 'Continue'. This causes a regeneration of the frameset, which will now show hyperlinks to the reviews of other reviewers.

Not all reviewers like to fill in review forms on-line. Therefore, a template is provided to the reviewers which they can use to fill in reviews off-line and send them by email. CyberChair checks the review forms for correctness and transforms them into a format that can be used for display in a browser.

To be able to protect the reviews from outsiders, each reviewer gets his or her own directory on the webserver. Access is controlled by a file called ".htaccess". This file contains the login names of those who are allowed to access the files in that directory. All review files a reviewer needs, such as the review form and other reviewers' reviews, are written in the reviewer's directory. Each time the reviewer clicks on a paper number in the top frame, CyberChair checks if the reviewer has already submitted his or her review. If this is the case, then the reviews of the other reviewers for this paper are copied into the reviewer's directory. This way, the reviewer can always look at the latest version of the reviews. When displaying a review, the date and time of the last update of the review is shown as extra check.

**RELATED WORK**

Although more conference management systems exist [11] and many conferences use electronic uploading and some kind of reviewing system, no such system was found in the literature.

A commonality between BSCW[2] and CyberChair is that both provide a shared workspace. BSCW could therefore be used to store abstracts, papers and reviews. It is not clear to us if BSCW could also be used to create a dynamic review system.

From personal communication with its creator we know that the conference on Object-Oriented Programming, Systems, Languages and Applications (OOPSLA) uses a system like CyberChair, which is implemented with so-called Java servlets.

The Artefact Framework [3] uses html frames like CyberChair. The authors claim that there is no provision for a server to send asynchronous updates. We showed that browsers of today are able to send requests without the user taking action.

The problematic use of groupware applications, as reported in several publications [7][10] is not affecting CyberChair. One reason is that users like the system. Other reasons are that authors are forced to use it to submit their paper. For reviewers, the features available to them are options. If a reviewer wishes, he or she may send reviews by email, and never look at his or her webpage. Further, a reason for easy acceptance by its users is that CyberChair is in fact merely an automation of a fairly simple, well known process.

In The Information Lens [8] the authors focus on sharing information among a group of people to solve problems. This problem is handled by CyberChair by the conflict detection mechanism and availability of other reviewers' reviews. In doing so, the PC can prepare themselves optimally for the PC meeting, thereby avoiding the problems mentioned by Stefik et al. [12], who provide groupware support for collaboration and problem solving *during* meetings.

A commonality between CyberChair and gIBIS [4] is that both systems provide means to make high-quality decisions by providing the group members' arguments to each other.

We hope, by describing CyberChair, to have answered A. Dix's questions about the usefulness of the web for CSCW [6].

**CONCLUSIONS**

CyberChair significantly reduces the workload of the PC, by doing all necessary bookkeeping. It eases the submission of author information, papers and reviews instantly in electronic form, using the web. The reviewers' reviews are compared to the other reviews to detect conflicts. Further, it allows reviewers to learn about each other's opinions, which provides means for good preparation for the PC meeting. CyberChair supports preparation of the proceedings by generating the table of contents and author index, based on the input of the authors.

More conference management systems exist [11], although we could not find publications about them in the litterature. As far as we know from the comments of its users, CyberChair is the only system that has implemented the patterns identified by Nierstrasz. The conflict detection and coloring scheme used to indicate the state of reviews has, as far as we know, not been used in other systems that handle review processes. The paper distribution problem addressed by CyberChair is using theory from another discipline: Mathematics.

CyberChair collects abstracts, papers and reviews and can therefore be regarded as a system that provides a shared workspace. It helps the PC to select the best papers for the conference, which makes it a group decision support sys-tem. Since the authors and the PC are working together towards the proceedings, it can also be seen as a collaborative authoring system.

As of this writing, CyberChair has been used successfully for 10 conferences or workshops, while currently half a dozen other conferences are planning to use it. A demonstration system of CyberChair is available at its homepage [5].